# Breathing k-Gap Events and ‘Instability on Instability’ in Nonlinear Photonic Time Crystals

Liang Zhang, Chenhao Pan, Yiming Pan*

State Key Laboratory of Quantum Functional Materials, School of Physical Science and Technology and Center for Transformative Science, ShanghaiTech University, Shanghai 200031, China

**Abstract**

Photonic time crystals (PTCs) host momentum bandgaps, or k-gaps, that enable parametric amplification and lasing of seeded fields. In nonlinear PTCs, Kerr saturation dynamically suppresses the exponential growth, reshaping k-gap amplification into an active, spatially homogeneous k-gap soliton train. Here, we show that a localized perturbation on this unstable background then nucleates a transient spatiotemporal excitation: the breathing k-gap event. Unlike Peregrine breathers emerging from modulational instability on a plane-wave background, this event extracts energy from competing host k-gap solitons and remains sustained by their interaction. We identify this process as an “instability on instability” mechanism intrinsic to nonlinear k-gap dynamics. The event is robust against noise and disorder, and can be deterministically reshaped into collective breathing patterns by periodic and phase-engineered seeding. These results establish k-gap engineering as a route to generating and controlling extreme spatiotemporal waves in photonic time-varying media.

Nonlinear wave systems support a class of extreme wave excitations that appear from a continuous-wave (CW) background, reach large peak amplitudes, and subsequently vanish without forming persistent localized states. Peregrine breathers [1], Rogue waves [2,3], and event solitons [4] are representative examples, typically associated with modulational instability (MI) while intrinsic spacetime localization. In contrast to long-lived, particle-like solitons, such excitations are better understood as transient nonlinear events rather than stationary wave states.

Photonic spacetime crystals (STCs) [5] provide a natural and tunable setting for event-like localization because simultaneous spatial and temporal modulation of the dielectric constant opens a mixed energy-momentum gap, or ωk-gap. In this full-gap regime, event solitons [4] are stabilized by the balance among the ω-gap-induced group-velocity dispersion, k-gap amplification, and nonlinearity-induced saturation, while topological events [6–8] arise from spacetime defects and are topologically protected by a winding number. Thus, known STC event formation relies on localized wavepackets embedded in a full ωk-gap, with stability supplied by either nonlinear balance or topological protection.

A simpler yet equally intriguing setting is the photonic time crystals (PTCs) [9], a spatially uniform medium with permittivity modulated only in time. Such modulation opens a momentum bandgap (k-gap), inside which modes acquire complex frequencies and undergo exponential amplification or attenuation through energy exchange with the modulated medium [10]. With Kerr nonlinearity, unstable k-gap modes can self-organize into superluminal k-gap solitons from a uniform seed [11], stabilized by the balance between k-gap gain and nonlinear saturation. However, unlike STCs, PTCs lack spatial modulation and a full ωk-gap, leaving it unclear whether purely time-varying media can support controllable nonlinear event-like excitations. This raises the central question addressed here: can k-gap physics alone generate and control spatiotemporal wave events?

Here, we report a spatiotemporal excitation in nonlinear PTCs: the breathing k-gap event. Unlike the Peregrine breather, which emerges from modulational instability on a stationary background, this event develops on an unstable k-gap soliton background. A local perturbation seeded onto this uniform soliton train aggressively breaks the nonlinear saturation, triggering energy extraction from both the host field and temporal pump to nucleate a breathing spatiotemporal burst. We identify this secondary instability on an already unstable background as an instability-on-instability mechanism. By varying the nonlinear strength and seed amplitude, we show that the event peak intensity, onset time, and spatial confinement can be controlled independently. We further show that periodic and phase-engineered seeding converts the competition between event breather and k-gap soliton into collective breathing patterns with tunable morphology and spatial organization. These results establish nonlinear k-gap engineering as a route to generating and controlling spatiotemporal wave events in time-varying media.

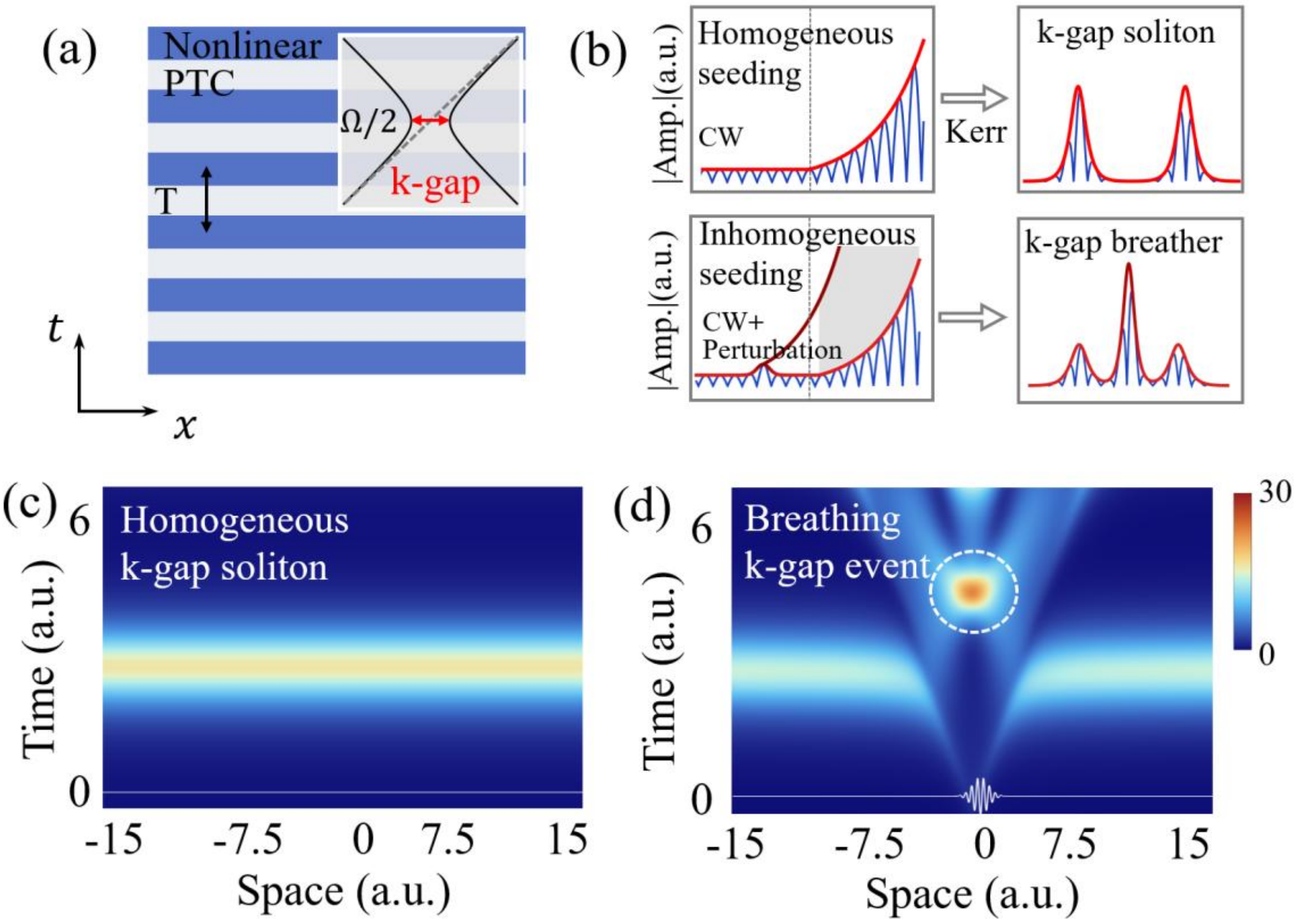

**Figure 1| From k-gap amplification to a breathing k-gap event in nonlinear photonic time crystals (PTCs).** (a) Schematic of a nonlinear PTC with temporally periodic permittivity (period $T = 2\pi/\Omega$). Inset: Floquet dispersion shows the time modulation opens a momentum gap (k-gap) at $\omega = \Omega/2$. (b) Mechanism of event breather formation: k-gap amplification of an in-gap continuous-wave seed is saturated by Kerr nonlinearity, yielding a spatially homogeneous k-gap soliton train in space. Adding a local perturbation then induce inhomogeneous amplification and nucleates a spatially localized breathing k-gap event. (c) Simulation results for homogeneous k-gap soliton background sustained by the balance between k-gap amplification and Kerr

nonlinearity. (d) Simulation results for spatiotemporal pattern of the breathing k-gap event seeded from a local perturbation onto the unstable background.

**Momentum k-gap engineering.** We consider a nonlinear photonic time crystal (PTC) [Fig. 1(a)], modeled as a spatially homogeneous medium with temporally periodic permittivity. Under the slowly varying envelope approximation, Maxwell's equations reduce to coupled-mode equations for the forward ($A_f$) and backward ($A_b$) propagating envelopes (see Methods):

$$\begin{aligned} \frac{i}{c}\frac{\partial A_f}{\partial t} + i\frac{\partial A_f}{\partial x} + \kappa A_b + \gamma\left(\left|A_f\right|^2 + 2|A_b|^2\right)A_f = 0, \\ -\frac{i}{c}\frac{\partial A_b}{\partial t} + i\frac{\partial A_b}{\partial x} + \kappa A_f + \gamma\left(|A_b|^2 + 2\left|A_f\right|^2\right)A_b = 0. \end{aligned} \tag{1}$$

Here $c = c_0/n_0$ denotes the light speed in the unmodulated background medium, with $n_0 = \sqrt{\epsilon_0}$ being the refractive index. The temporal modulation gives the coupling $\kappa = \delta\Omega/8c$, with modulation depth $\delta \ll 1$ (typically up to 0.2) and frequency $\Omega = 2\pi/\mathrm{T}$. At visible frequencies, this corresponds to femtosecond-scale modulation periods. The Kerr nonlinearity is characterized by $\gamma = \beta\Omega/4c$, where $\beta = 3\chi_3/\epsilon_0\epsilon_r^3$. Equation (1) captures the competition between mode coupling and nonlinear phase modulation [12], and serves as the minimal model for k-gap formation, homogeneous k-gap solitons, and breathing k-gap events.

Equation (1) can be recast as a nonlinear Dirac-like form, defining $\psi = \left(A_f, A_b\right)^T$, we have $\left(\frac{1}{c}\left(i\frac{\partial}{\partial t}\right) + \sigma_z\left(i\frac{\partial}{\partial x}\right)\right)\psi + i\kappa\sigma_y\psi + \gamma\sigma_z\mathcal{L}_{\mathrm{NL}}[\psi]\psi = 0$, where $\psi$ is the two-component spinor field and $\sigma_{y,z}$ are the Pauli matrices. The nonlinearity term is defined as $\mathcal{L}_{\mathrm{NL}}[\psi]\psi = \psi^\dagger(3\sigma_0 - \sigma_z)\psi/2$. In this presentation, the temporal coupling $\kappa$ acts as an imaginary Dirac mass, opening a momentum k-gap rather than the frequency gap associated with a real Dirac mass. Setting $\gamma = 0$ temporarily, using $\psi = \chi e^{i(kx-\omega t)}$, where $\chi$ is a two-component spinor, gives the dispersion relation $\omega^2 = c^2(k^2 - \kappa^2)$. Therefore, modes within the interval $k < |\kappa|$ have complex frequencies, and the interval is defined a k-gap of width $2\kappa = \delta\Omega/4c$ (Fig. 1a), directly controlled by the temporal modulation.

The effective parameters in Eq. (1) can be engineered in time-varying platforms including microwave circuits [7,13,14], acoustics [15,16], and metamaterials [17–19].

We therefore use the dimensionless form of the model to focus on the shared k-gap physics and nonlinear wave dynamics across these platforms.

We first examine the nonlinear evolution of a linearly unstable k-gap mode. Simulations are performed in dimensionless units with $c = 1, \kappa = 1$ in a window $x \in [-L, L]$, with $L = 15$. The initial state is a weak continuous-wave (CW) seed whose carrier $(\omega_0, k_0)$ lies inside the k-gap. For weak Kerr nonlinearity $\gamma = 0.05$, the early-stage dynamics are governed by exponential k-gap amplification. As the intensity increases, nonlinear phase modulation saturates the gain and prevents k-gap amplification (Fig. 1c). The field then enters a recurrent growth-saturation-relaxation cycle, forming a temporally periodic and spatially homogeneous k-gap soliton train. This train provides the active unstable background field on which breathing k-gap event are later seeded.

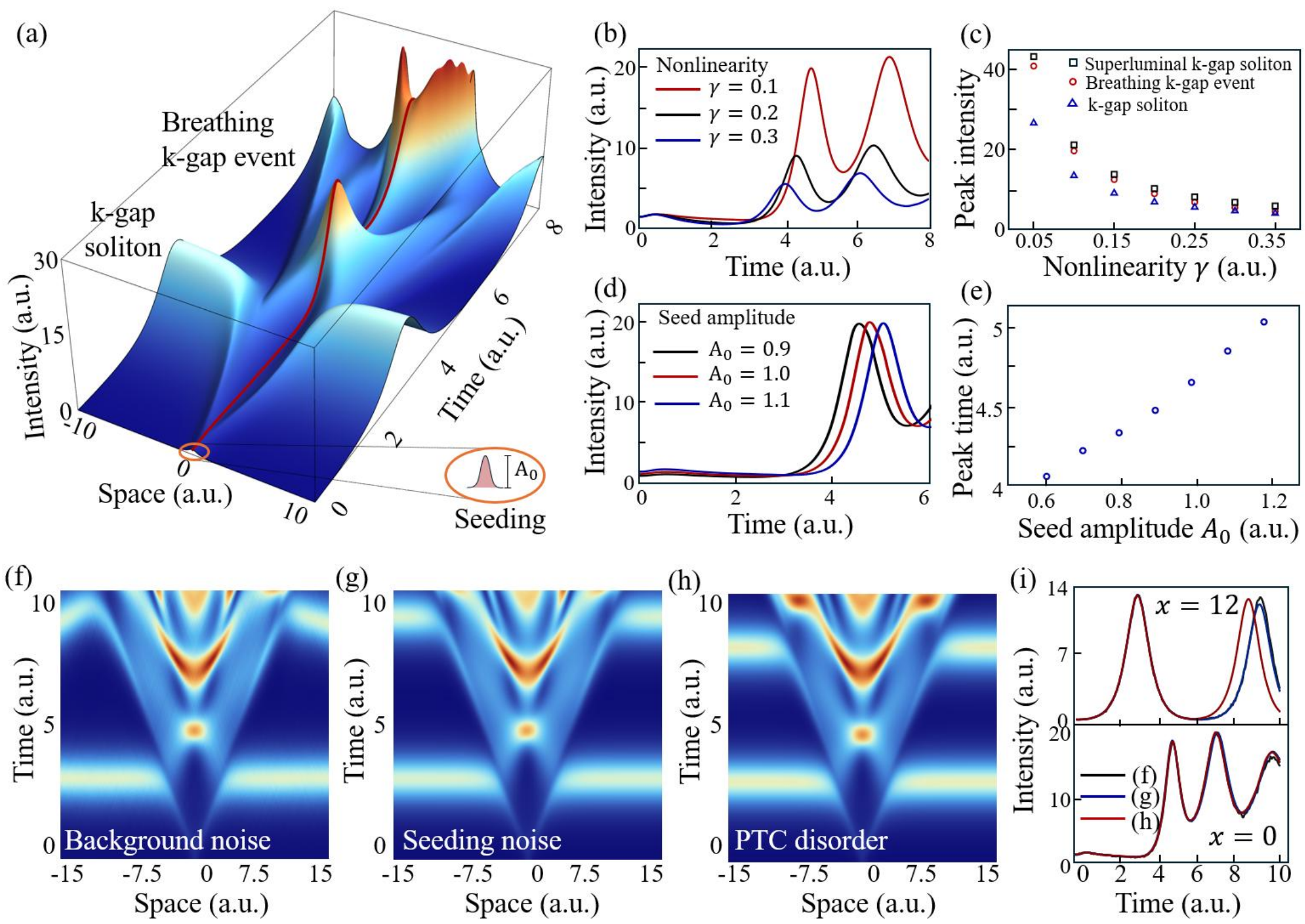

**Figure 2| Parameter control and robustness of breathing k-gap events.** (a) Spatiotemporal intensity landscape showing the evolution of an event breather. (b) Temporal intensity profiles at the nucleation sites for different nonlinearity strengths $\gamma$, showing coexistence of the breathing event and the k-gap soliton. (c) Peak intensities of the k-gap soliton (squares), breathing k-gap event (circles), and background soliton (triangles) as a function of $\gamma$. (d) Temporal intensity traces for different seed amplitudes $A_0$, showing that $A_0$ mainly shifts the event onset time while leaving the

peak profile remains nearly unchanged. (e) Event peak time $t_p$ as a function of $A_0$, demonstrating a deterministic control of the onset time. (f)-(h) Spatiotemporal intensity maps under background noise, seeding noise and temporal disorder of the PTC, respectively. (i) Temporal intensity traces extracted at $x = 12$ for the k-gap soliton train (top) and at $x = 0$ for the breathing k-gap event (bottom) in (f)-(h).

**Formation of the breathing k-gap event.** We next seed the unstable k-gap soliton background with a localized perturbation at $t = 0$ of the form $0.25\left(1 + A_0(-\frac{4(1-2ix)e^{-x^2/10}}{1+4x^2})\right)$, where $A_0$ sets the seed amplitude. Using Eq. (2), the perturbation does not merely nucleate another k-gap soliton; instead, it disrupts the background's homogeneity of the saturated train and drives the formation of a spatiotemporally localized burst, which we identify as the breathing k-gap event, because it displays three simultaneous signatures: local enhancement at the seeded site, depletion of the surrounding host train, and a finite-time breathing cycle before reorganization into a propagating ridge [Fig. 1(d)]. Throughout this process, the signal remains bounded by the light cone, preserving causality [20].

Figure 2a resolves the breathing k-gap event as a multi-stage dynamical process. A localized perturbation embedded in the extended soliton train first grows into a local spatialtemporal peak under k-gap amplification. The surrounding train is depleted as the peaks intensifies, indicating that the event is dynamically coupled to the active nonlinear background rather than from a passive CW pedestal. The peak then forms a sharply localized burst, undergoes a breathing cycle of broadening and recompressing, and finally reorganized into a propagating ridge associated with a superluminal k-gap soliton.

To assess the role of nonlinearity, we fix the initial perturbation profile and vary the Kerr coefficient $\gamma$. As $\gamma$ is reduced, the same k-gap gain is arrested at a larger field amplitude, allowing both the breathing event and the subsequent propagating ridge to reach higher intensity, as shown in Fig. 2(b). The extracted peak intensities in Fig. 2(c) show that the event peak, the superluminal soliton and the background soliton train all decrease with increasing $\gamma$. Among them, the breathing event exhibits the strongest dependence on $\gamma$. This tunable peak response contrasts with the Peregrine breather, whose peak-to-background ratio is fixed, and shows that the breathing k-gap event is continuously controlled by the nonlinear strength.

By contrast, the seed amplitude $A_0$ primarily controls the timing of the event. At fixed $\gamma$, increasing $A_0$ shifts the burst to later times while leaving the event peak intensity and spatiotemporal profile nearly unchanged [Fig. 2(d)]. This delayed peak indicates that the seed does not simply undergo threshold-like linear amplification; rather, it modifies the local k-gap amplification–saturation balance, so that the observed maximum is set by the subsequent breathing and recompression dynamics.The extracted peak time is controlled by $A_0$ [Fig. 2(e)]. Thus, $\gamma$ mainly sets how large the event becomes, whereas $A_0$ tunes when it appears [Fig. 2(e)].

We further examine three representative perturbations: background noise, fluctuations in the localized seed, and temporal disorder of the PTC coupling. These tests probe different failure channels. Background noise tests whether the event survives fluctuations of the host train; seed noise tests sensitivity to imperfect excitation; and temporal disorder tests whether strictly periodic modulation is required. In all three cases, the breathing event remains identifiable, although different parts of the dynamics are modified, as shown in Fig. 2(f)-2(i). Background noise roughens both the train and event profile, seed fluctuations break the event spatial symmetry without destroying its localized contour, and temporal disorder modifies the period of the host soliton train and affects the subsequent superluminal k-gap soliton, but the breathing event itself remains stable. These results show that the breathing k-gap event is not a fine-tuned artifact of an ideal input.

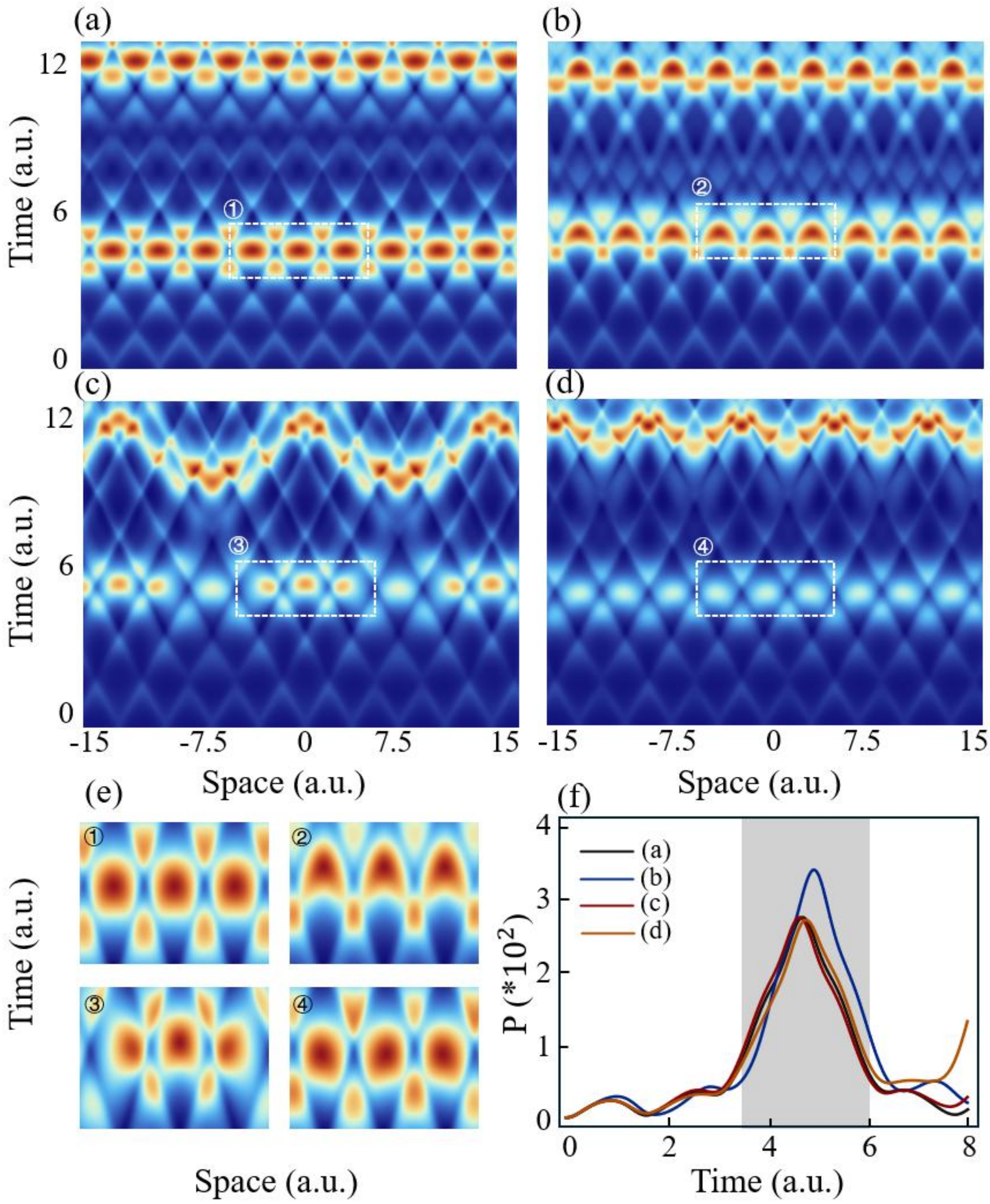

**Figure 3| Phase-engineered collective patterns of breathing k-gap events.** (a) spatiotemporal intensity generated by a spatially periodic perturbation seed. (b) Pattern obtained by adding a uniform linear phase gradient to the periodic seed. (c) Pattern generated by a weak long-period sinusoidal phase modulation, forming a grouped event array. (d) Pattern generated by a half-resonant sinusoidal phase modulation, yielding a regular event sequence. (e) Enlarged views of the dashed regions in (a-d), showing the corresponding event super-cell structures. (f) Spatially integrated intensity $P(t)$ for

the four cases in (a-d). The shaded window marks the temporal interval enlarged in (e) and shows how phase-engineered seeding modifies these transient dynamics.

**Phase-engineered breathing patterns.** To explore the collective dynamics of these event excitations, we replace the isolated seed with a spatially periodic perturbation. Maintaining the previous parameters ($\kappa = 1, c = 1,$ and $L = 15$) but employing a stronger nonlinearity ($\gamma = 0.1$), we initialize the field with the periodic seed $I(x) = 0.1\rho + \frac{(1+[2(1-2a_0)\cos B_0 x + iB_0 \sin B_0 x])}{\sqrt{2a_0} - \cos B_0 x}$, with $B_0 = \sqrt{8a_0(2a_0-1)}$, $\rho = \sqrt{2/3\gamma}$ and $a_0 = 0.8$. This profile represents a k-gap soliton background dressed by a periodic perturbation with spatial period $2\pi/B_0$. Evolving Eq. (1) from this initial state generates a regular breathing pattern rather than an isolated event [Fig. 3(a)]. Physically, this pattern can be understood as a periodic train of localized breathing k-gap events induced on the extended host background. Each event exchanges energy with the host soliton train while simultaneously competing and interfering with neighboring events, which collectively gives rise to the pronounced spatial periodicity of the pattern.

We then further tailor breathing dynamics through phase engineering of the seed. First, imposing a linear phase gradient, $I_1(x) = I(x)e^{iqx}, q = 2\pi/L,$ preserves the overall periodic arrangement but strongly reshapes the local breathing event, transforming the originally regular profile into a crescent-like structure [Fig. 3(b)]. Next, we apply a periodic phase modulation of the form $I_2(x) = I(x)e^{i\phi_2(x)}, \phi_2(x) = A_p \cos(Q_p x)$. For $A_p = 0.2\pi$ and $Q_p = B_0/4$, the local event shape is further modified and the spatial periodicity of the full breathing pattern is evidently reorganized [Fig. 3(c)]. Increasing the modulation wave number to $A_p = 0.1\pi$ and $Q_p = B_0/2$ produces another distinct breathing structure with a different periodic arrangement [Fig. 3(d)]. These differences are shown more clearly in the enlarged views in Fig. 3(e). Therefore, the imposed seed phase not only reshapes individual breathing events, but also provides an effective handle for tuning the collective structure and spatial periodicity of the entire breathing pattern. Moreover, the differences among these phase-engineered cases become increasingly pronounced at later times, showing that seed-phase modulation governs not only the initial morphology, but also the subsequent collective spatiotemporal evolution.

To quantify the global evolution, we further compute the spatially integrated intensity $P(t) = \int_{-L/2}^{L/2} (\left|A_f(x,t)\right|^2 + |A_b(x,t)|^2) dx,$ as shown in Fig. 3(f). The shaded region marks the main temporal window displayed in Fig. 3(e), where the breathing pattern is most prominent. Compared with the unmodulated case, the linear-phase seed delays the emergence of the breathing pattern, indicating that the localized breathing dynamics set in at a later time. At the same time, this delay allows the system to accumulate a higher total intensity before the breathing burst is fully developed, leading to the largest peak in $P(t)$, whereas the periodic phase modulation mainly reshapes the breathing pattern without noticeably changing the intensity of the breathing event itself. Therefore, the imposed seed phase does not merely deform individual breathing events, but also provides an effective handle for tuning the collective structure and spatial periodicity of the entire breathing pattern.

**Further discussion**

The breathing k-gap event identified here is best understood as a finite time localized excitation on an active k-gap soliton train, rather than as a simple variant of a conventional breather. Its key feature is that the background itself is generated by k-gap amplification and nonlinear saturation; the localized seed then induces a secondary localization process on this active and gain saturated host field. Unlike conservative breather solutions, whose background are usually prescribed by the same nonlinear dynamics, the present event is tied to an externally pumped gain-saturation cycle and can reorganize into a propagating k-gap soliton after the burst. We have focused on a one-dimensional scalar nonlinear PTC, and the robustness discussed here refers to the tested background noise, seed disorder, and temporal-modulation disorder, rather than to topological protection or universal stability. A more systematic stability analysis and higher-dimensional generalizations remain open problems.

Experimentally, time-modulated microwave transmission lines and nonlinear circuit metamaterials appear to be the most realistic near-term platforms. In such platforms, the modulation depth sets the effective coupling $\kappa$, the nonlinear loading controls $\gamma$, and the injected waveform determines the seed amplitude and phase. However, observing the effect would require not only a measurable k-gap and nonlinear saturation, but also sufficiently low loss, controlled seeding, and suppression of unwanted nonlinear distortion. A practical first step would be to observe the sequence of k-gap amplification, nonlinear saturation into a recurrent host train, and seed-induced localized breathing.

## Conclusion

We reported that nonlinear photonic time crystals host a spatiotemporal event excitation: the breathing k-gap event. We find that temporal modulation dictates a primary momentum k-gap instability, which Kerr nonlinearity reshapes into a self-organized soliton train. Superimposing a secondary modulational instability onto this active soliton background inevitably triggers the nucleation of transient, intensely localized spatiotemporal bursts. By aggressively extracting energy from the unstable host field, these bursts epitomize an underlying "instability on instability" mechanism. We further demonstrate that seed and phase engineering provide deterministic control over both isolated events and collective breathing patterns. These findings open a route to engineering spatiotemporally localized structures through the interplay of k-gap physics and nonlinearity, and suggest an underlying mechanism for exploring and controlling event dynamics in time-varying media.

## Declarations

### Availability of data and materials

The data that support the findings of this study are available from the corresponding authors upon reasonable request.

### Competing interests

The authors declare that they have no competing interests.

### Funding

This work was supported by the NSFC (No. 2023X0201-417-03) and the start-up funding from ShanghaiTech University.

### Authors' contributions

**Liang Zhang:** Conceptualization, methodology, numerical simulations, formal analysis, visualization, writing – original draft. **Chenhao Pan:** Discussion and manuscript revision. **Yiming Pan:** Conceptualization, supervision, funding acquisition, writing – review & editing. All authors read and approved the final manuscript.

*yiming.pan@shanghaitech.edu.cn

## Acknowledgements

Not applicable.

## Authors' information

Not applicable.

**Methods.**

We derive our theoretical framework from Maxwell's equations for a non-magnetic, source-free ($\rho = 0, \boldsymbol{J} = 0$) dielectric medium with an instantaneous nonlinearity. Applying the curl operator to Faraday's law and substituting Ampere's law yields the fundamental wave equation:

$$\frac{\partial^2 \boldsymbol{D}}{\partial t^2} = -\frac{1}{\mu_0} \nabla \times (\nabla \times \boldsymbol{E}). \qquad (M1)$$

Assuming an optical Kerr effect for definiteness, the constitutive relation in a 1D isotropic medium is given by $\boldsymbol{D} = \epsilon(t, |\boldsymbol{E}|^2)\boldsymbol{E} = \epsilon_0[\epsilon_1(t) + \chi_3|\boldsymbol{E}|^2]\boldsymbol{E}$ with $\chi_3$ being the Kerr coefficient. The linear dielectric constant is chosen to be spatially homogeneous but periodically modulated in time

$$\epsilon_1(t) = \epsilon_r(1 + \delta \cos \Omega t), \qquad (M2)$$

where $\epsilon_r$ is the static relative permittivity, $\delta \ll 1$ is the dimensionless modulation depth (e.g., up to 0.3) and $\Omega = 2\pi/T$ the modulation frequency. For physical context, driving a PTC in the visible-light regime (e.g., an 800 nm carrier wavelength) necessitates an ultrafast temporal modulation period of $T$=2.7 fs.

To proceed, we invert the constitutive relation to expand the electric field in terms of the displacement field. Dropping the vector notation for the 1D case, we obtain $\boldsymbol{E} = \frac{\boldsymbol{D}}{\epsilon_0\epsilon_1} - \frac{\chi_3 \boldsymbol{D}^3}{\epsilon_0^2\epsilon_1^4}$. Crucially, we formulate the nonlinear wave equation strictly in terms of $\boldsymbol{D}$ rather than $\boldsymbol{E}$ [21]. In a temporal crystal, the displacement field $\boldsymbol{D}$ guarantees rigorous continuity across temporal boundaries, a fundamental condition not necessarily satisfied by $\boldsymbol{E}$ . Substituting this expansion into Eq. (M1) yields:

$$\frac{\partial^2 D}{\partial t^2} = \frac{1}{\mu_0} \frac{\partial^2}{\partial x^2} \left( \frac{D}{\epsilon_0\epsilon_1} - \frac{\chi_3 D^3}{\epsilon_0^2\epsilon_1^4} \right) \qquad (M3)$$

Given the perturbative modulation depth ($\delta \ll 1$), we safely neglect its dynamic contribution to the higher-order Kerr term. Retaining the dominant self-phase modulation contribution, the wave equation simplifies to:

$$\frac{1}{c^2} \frac{\partial^2 D}{\partial t^2} = (1 - \delta \cos \Omega t) \frac{\partial^2 D}{\partial x^2} - \beta |D|^2 \frac{\partial^2 D}{\partial x^2} \qquad (M4)$$

where $c = c_0/\sqrt{\epsilon_r} = c_0/n_0$ is the background speed of light and $\beta = 3\chi_3/\epsilon_0\epsilon_r^3$ dictates the effective nonlinear response.

To capture the $k$-gap dynamics, we employ a coupled-mode framework based on the Floquet-Bloch theorem, constructing the field as a superposition of forward- and backward-propagating envelopes:

$$D(x,t) = A_f(x,t)e^{ikx-i\omega t} + \mathrm{A}_b(x,t)e^{\mathrm{ikx}+i\omega t} + c.c., \qquad (M5)$$

where $A_f$ and $A_b$ are the slowly varying amplitudes of forward- and backward-propagating modes, scattering with each other via modulation-induced time reflection. To maximize the scattering, the carrier frequency and momentum are phase-matched to the modulation such that $\omega = \Omega/2$ and $k = \Omega/2c$. Substituting the ansatz into Eq. (M4) and invoking the slowly-varying envelope approximation, we integrate out the fast-oscillating terms to obtain the governing nonlinear coupled-mode equations:

$$\begin{aligned} \frac{i}{c}\frac{\partial A_f}{\partial t} + i\frac{\partial A_f}{\partial x} + \kappa A_b + \gamma\left(\left|A_f\right|^2 + 2|A_b|^2\right)A_f &= 0 \\ -\frac{i}{c}\frac{\partial A_b}{\partial t} + i\frac{\partial A_b}{\partial x} + \kappa A_f + \gamma\left(|A_b|^2 + 2\left|A_f\right|^2\right)A_b &= 0 \end{aligned} \qquad (M6)$$

where $\kappa = \delta\Omega/8c$ defines the temporal coupling coefficient and $\gamma = \beta\Omega/4c$ represents the effective Kerr coefficient.